\documentclass[authoryear,preprint,review,12pt]{elsarticle}

\usepackage{amssymb}
\def\astrobj#1{#1}

\journal{New Astronomy}

\begin{document}

\begin{frontmatter}

\title{Marginally low mass ratio close binary system V1191 Cyg}

\author[ege]{B. Ula{\c s}}
\author[ege,iyt]{B. ̃Kalomeni}
\author[ege]{V. Keskin}
\author[ege]{O. K\"ose}
\author[ege]{K. Yakut}
\address[ege]{Department of Astronomy and Space Sciences, University of Ege, 35100, {\.I}zmir, Turkey}
\address[iyt]{Department of Physics, {\.I}zmir Institute of Technology, 35430, {\.I}zmir, Turkey}
%\cortext[cr]{Corresponding author}

\begin{abstract}
In this study, we present photometric and spectroscopic variations
of the extremely small mass ratio ($q\simeq 0.1$) late-type contact binary system \astrobj{V1191 Cyg}.
The parameters for the hot and cooler companions have been determined
as $M_\textrm{h}$ = 0.13 (1) $M_{\odot}$, $M_\textrm{c}$ = 1.29 (8) $M_{\odot}$, $R_\textrm{h}$ = 0.52 (15) $R_{\odot}$,
$R_\textrm{c}$ = 1.31 (18) $R_{\odot}$, $L_\textrm{h}$ = 0.46 (25) $L_{\odot}$, $L_\textrm{c}$ = 2.71 (80) $L_{\odot}$,
the separation of the components is $a$= 2.20(8) $R_{\odot}$  and the distance of the system is
estimated as 278(31) pc. Analyses of the times of minima indicates a period increase
of  $\frac{dP}{dt}=1.3(1)\times 10^{-6}$ days/yr that reveals a very high mass transfer rate of $\frac{dM}{dt}=2.0(4)\times 10^{-7}$$M_{\odot}$/yr
from the less massive component to the more massive one.
New observations show that the depths of the minima of the light curve have been interchanged.
\end{abstract}


\end{frontmatter}

% \linenumbers

\section{Introduction}
\label{intro}
Modelling the evolution of late-type contact binary systems (LTCBs) is highly complicated.
Yakut \& Eggleton (2005) (hereafter YE05) modelled late-type close binary systems by assuming conservative and non-conservative cases.
The authors proposed that a detached configuration can evolve into a semi-detached and then to a
contact configuration. The mass ratio of the components reverses through evolution.
Therefore, currently observed low mass star could be initially the more massive one in a binary system.
The mass ratio of the components, which is related to the
angular momentum loss and mass transfer, is one of the crucial parameter in  the evolution of close binary systems.
Apart from \astrobj{V1191 Cyg}, there exist rare well-known LTCBs with a small mass ratio (e.g. \astrobj{SX Crv},
\astrobj{V870 Ara}, \astrobj{FG Hya}, \astrobj{$\epsilon$ CrA}, \astrobj{CK Boo}, \astrobj{FP Boo}, \astrobj{DN Boo},
\astrobj{AH Cnc}; see YE05, {\c S}enavc{\i} et al. 2008, Yakut et al. 2009).

\astrobj{\astrobj{V1191 Cyg}} was classified as a W UMa type system by Mayer in 1965. Many years later, the first
photometric study of the system was made by Pribulla et al. (2005a) and they derived the geometric elements of the
binary system without a spectroscopic study.
The authors gave the mass ratio of the components as 0.09 and reported orbital period increase of
$4.2\times10^{-6}$ yr$^{-1}$. Rucinski et al. (2008) studied the radial velocity curve
of \astrobj{V1191 Cyg} and determined the spectroscopic mass ratio, orbital period,
radial velocity amplitudes, and the velocities of the center of mass of the system.
However, the authors also mentioned that the system is indeed a difficult spectroscopic target
because of its faintness.

\section{New Observations}

\begin{figure*}
\includegraphics[height=80mm]{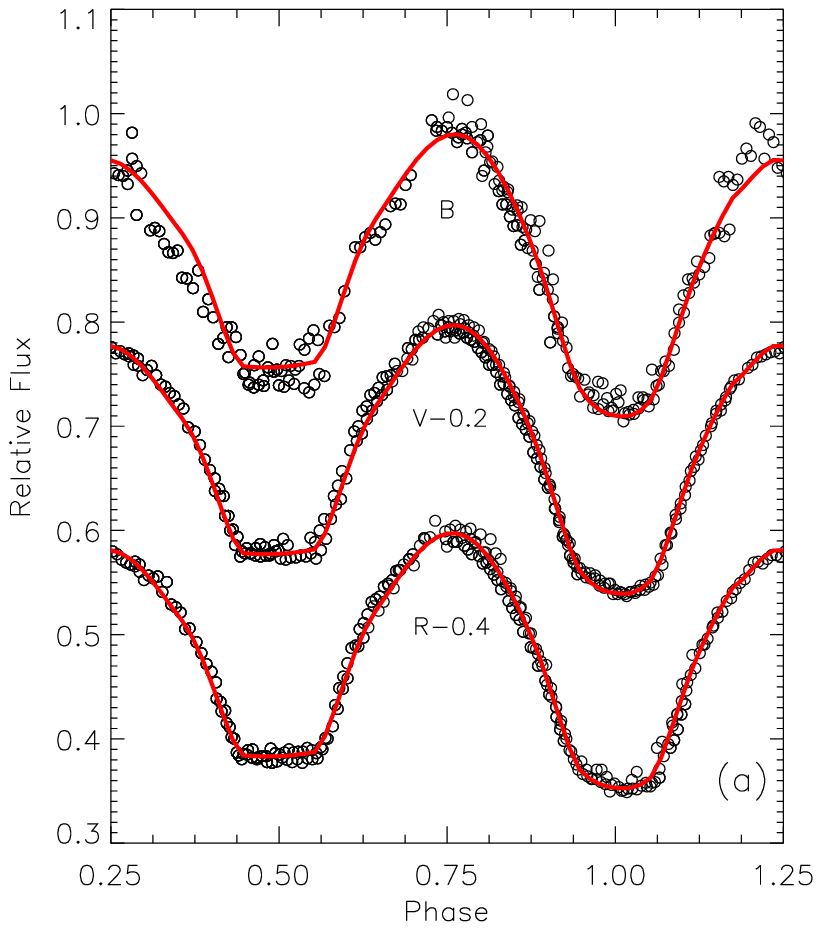}\vspace{-0.5cm}\\
\includegraphics[height=50mm]{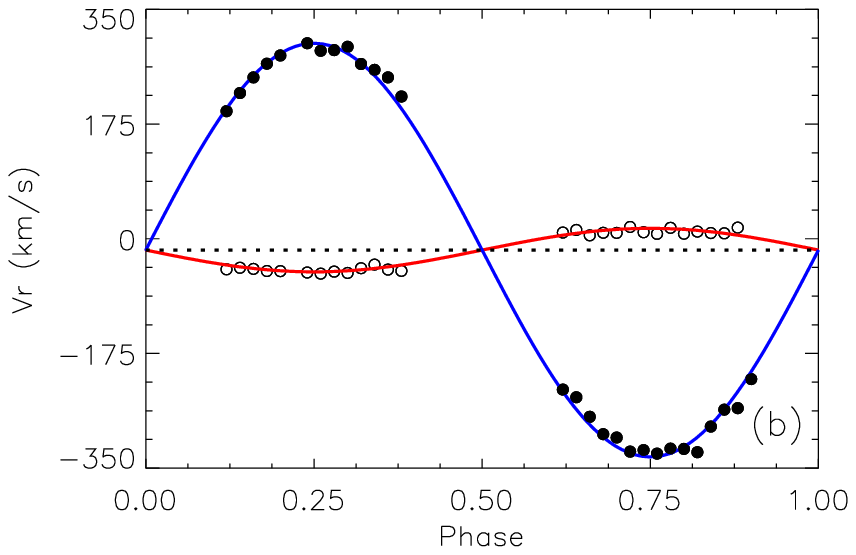}
\caption{(a) The observed and computed (solid line) light and (b) radial velocity curves of \astrobj{V1191 Cyg}.
Light curves in V and R bands are moved by a value of -0.20 and -0.40, respectively, for a good visibility.
Observed radial velocities are obtained from Rucinski et al. (2008).
The continous lines are estimated from simultaneous solution.}\label{Fig:V1191Cyg:LC}
\end{figure*}

The observations carried out with the 40-cm telescope at Ege University Observatory.
This telescope equipped with the Apogee CCD camera. The light curve of the system and its times of
minima light are obtained in the $B$, $V$, and $R$ bands in three nights in 2008 (Fig.~\ref{Fig:V1191Cyg:LC}a).
Comparison and check stars were selected as GSC 3159-1409 and GSC 3159-1701, respectively.
Because of the suspected case of GSC 3159-1701, we also chose an additional check star,
GSC 3159-1593, to examine any variability of the comparison star. 237, 274 and 280 points were obtained
in $B$, $V$, and $R$ bands, respectively.
The data are reduced by using IRAF (DIGIPHOT/APPHOT) packages. Standard deviations of observations
are estimated for B, V, and R bands as $0.002$, $0.001$, and $0.001$, respectively.
In Table~\ref{tablc} we list all the observed data.

Due to stellar activity and/or inhomogeneous stellar surface,
the light curve variation of the system shows maxima with different amplitudes (Fig.~\ref{Fig:V1191Cyg:LC}) known as O'Connell effect.
We considered this effect in our light curve analysis and further discussed it in Section~3.
The depths of the primary minima are $0^{m}.34$, $0^{m}.31$ and $0^{m}.30$ in $B$, $V$ and $R$ filters, respectively.
The secondary minima are shallow by  $0^{m}.060$ in $B$,  $0^{m}.055$ in $V$, and  $0^{m}.040$ in $R$ band.
Duration of the primary and secondary minima are $3^{h}.1$ and $3^{h}.2$, respectively.
The comparison of light curves presented in this study with earlier ones shows that
the depth and the shape of the light curve have been changed. We will discuss this in the last section.

We obtained four new minima times throughout these observations. They are given with those published
in Table~\ref{tab:V1191Cyg:mintimes} with their errors.
Using Table~\ref{tab:V1191Cyg:mintimes} we derived a new linear ephemeris (Eq.~\ref{eq:v1191cyg:1})
and used it for analyses.

\begin{equation}
HJD\,MinI = 24\,54635.5231(7)+0.3133896 (3)\times E \label{eq:v1191cyg:1}
\end{equation}

\begin{table}
\caption{Observational points of \astrobj{V1191 Cyg}. Heliocentric Julian Date, phase, and magnitudes (B, V, and R) are listed. Table~\ref{tablc} is given in its entirety in the electronic edition of
 this paper. A portion of it is shown here for guidance regarding its form and content.}
\label{tablc}
\begin{tabular}{llll}
\hline
\hline
HJD	        &	Phase	&$\Delta$m & Filter \\	
2454600+    &	 	    & mag       &   \\	
\hline
30.3282	&	0.425	&	0.047	&	B	\\
30.3304	&	0.430	&	0.041	&	B	\\
30.3325	&	0.437	&	0.080	&	B	\\
30.3346	&	0.444	&	0.078	&	B	\\
30.3367	&	0.451	&	0.088	&	B	\\
30.3388	&	0.458	&	0.095	&	B	\\
30.3410	&	0.464	&	0.086	&	B	\\
30.3431	&	0.471	&	0.080	&	B	\\
\hline
\end{tabular}
\end{table}

\section{Eclipse Timings and Period Study}
Orbital period study of the system was made previously by Pribulla et al. (2005a).
The authors reported very fast period increase of $\Delta P / P = 4.22\times10^{-6}$ yr$^{-1}$.
Recent observations show that the deeper primary is the transit eclipse.
Therefore, we define the transit minimum as a primary minimum (I) and  the occultation as a secondary minimum (II).
We present minima times obtained in this study together with those of earlier studies in Table~\ref{tab:V1191Cyg:mintimes}.

We solved the $O-C$ curve to find a parabolic variation that can be described as a mass transfer
from the less massive companion to the more massive one. A total of 35 data obtained by
photometric/CCD observations are used to study the period variation of the system.
The weighted least-squares method is used in order to determine the parameters of the upward parabolic variation.
The residuals ($\Delta T_I$) indicate a quadratic solution. In order to estimate the light elements
given in Eq. \ref{eq:v1191cyg:2} differential correction method is used.
Using a weighted least squares solution we obtain the parameters as:

\begin{equation}\label{eq:v1191cyg:2}
\begin{array}{c}
\rm{HJD\,MinI}  = 24\,52548.5135(1)  \\ + 0.31338491(2)\times E
+5.5(3)\times 10^{-10}\times E^2.
\end{array}
\end{equation}

\begin{table}
\begin{center}
\caption{The times of minima light of  V1191 Cyg.}\label{tab:V1191Cyg:mintimes}
\begin{tabular}{llll}
\hline
HJD* Min &     Ref &   HJD Min &     Ref \\
\hline
49587.3879	&     1       &       52905.305       &   6	  \\
49599.4536	&     1       &       52905.4591      &   6	  \\
49608.384	&     1       &       53156.4815      &   6	  \\
49619.3527	&     1       &       53512.4895      &   6	  \\
49619.5106	&     1       &       53258.4871      &   7	  \\
49621.3904	&     1       &       53340.2859      &   7	  \\
49644.2651	&     2       &       53612.4622      &   8	  \\
49644.4228	&     2       &       53685.642       &   9	  \\
50672.455	&     3       &       53915.5113      &   10	  \\
52413.445	&     4       &       53921.4649      &   10	  \\
52445.4067	&     4       &       53934.4683      &   10	  \\
52456.3748	&     4       &       54025.3548      &   11	  \\
52465.4622	&     4       &       54049.6383      &   12	  \\
52528.3033	&     4       &       54630.3527(3)   &   13	  \\
52548.3544	&     4       &       54635.3660(2)   &   13	  \\
52901.5459	&     5       &       54635.5230(5)   &   13	  \\
52902.3257	&     6       &       54636.4643(2)   &   13	  \\
52902.4848	&     6       &              	     &         \\
\hline
\end{tabular}
\end{center}
\scriptsize {References for Table~\ref{tab:V1191Cyg:mintimes}.
1	-	Agerer \& Hubscher (1995)	,
2	-	Agerer \& Hubscher (1996)	,
3	-	Agerer \& Hubscher (1998)	,
4	-	Pribulla et al. (2002)	,
5	-	Hubscher (2005)	,
6	-	Pribulla et al. (2005b)	,
7	-	Hubscher et al. (2005)	,
8	-	Hubscher et al. (2006)	,
9	-	Nelson (2006)	,
10	-	Parimucha et al. (2007)	,
11	-	Hubscher (2007)	,
12	-	Nelson (2007)	,
13	-	present study.}
\end{table}

\begin{figure*}
\includegraphics[height=110mm]{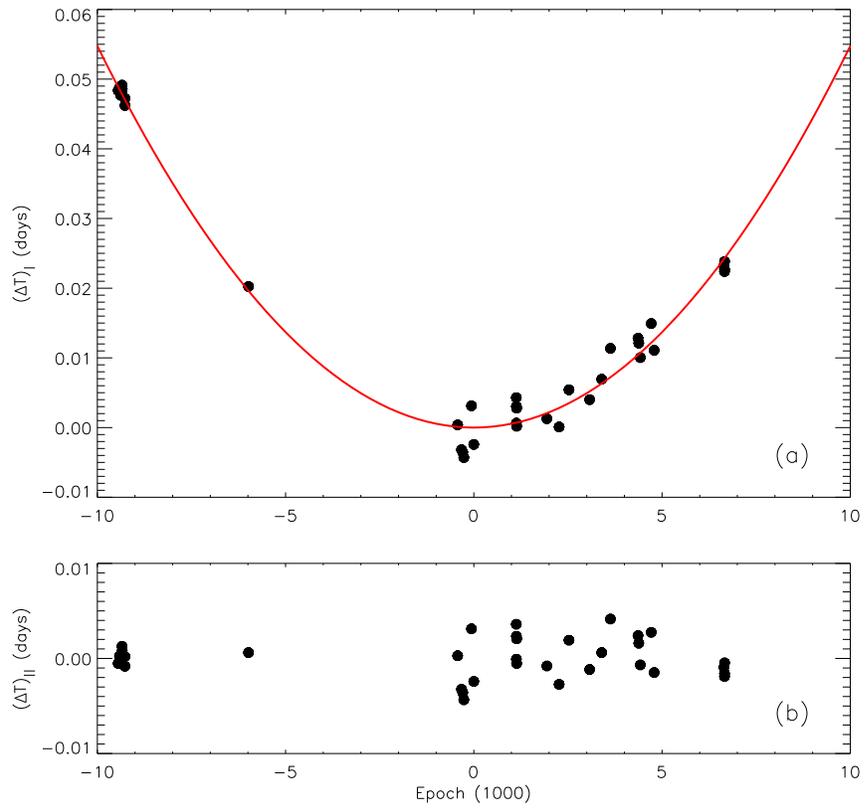} \caption{ (a) Residuals for the times of minimum
light of \astrobj{V1191 Cyg}. The solid line is obtained with the quadratic
terms in ephemeris (Eq.~\ref{eq:v1191cyg:2}). (b) The difference between the
observations and the quadratic ephemeris.}\label{Fig:V1191Cyg:OC}
\end{figure*}

\section{Simultaneous solutions of light and radial velocity curves}

The only photometric light curve study of the system was published by Pribulla (2005a)
 who applied the \textsc{roche} code to the $BVR$ data of two nights taken at
 the Stara Lesna Observatory. Pribulla first performed a grid search to the data to evaluate
 the mass ratio and resulted the analysis with geometric parameters $i=80^{\rm o}.4$, $q=0.094$, and $f=0.46$.
 In addition, Pribulla also mentioned about the possibility of the presence of a third body.
 The radial velocity of the system was studied by Rucinski et al. (2008). Spectroscopic orbital elements was given
 as $V_0 = -16.82$ km/s, $q=0.107$, $K_1 = 33.68$ km/s, and  $K_2 = 315.52$ km/s (ibid). Rucinski et al. (2008) reported
 smaller mass ratio for the system than an ordinary W-type W UMa system.

New light curves and radial velocity curves of Rucinski et al. (2008) simultaneously
have been solved by using the program {\sc phoebe} (Pr\~{s}a \& Zwitter 2005),
which uses the WD code (Wilson \& Devinney 1971). In the solution we used weighted light
curves which are constructed according to the standard deviation of each filter.
The effective temperature of the cooler star is chosen according to the spectroscopic study of Rucinski et al. (2008) in which its spectral type was given as F6V. The albedos,
 $A_\textrm{h}$ and $A_\textrm{c}$ are adopted from Rucinski (1969), and the
 values of the gravity darkening coefficients, $g_\textrm{h}$ and $g_\textrm{c}$, are taken from Lucy (1967).
 The logarithmic limb-darkening law are used with coefficients adopted from van Hamme (1993)
 for a solar composition star (Table~\ref{tab:V1191:lc}). The adjustable photometric parameters are
 orbital inclination, \textit{i}, surface potential, $\Omega_h = \Omega_c = \Omega$, temperature
 of the hot component, $T_\textrm{h}$, luminosity, $L_\textrm{h}$, mass ratio, $q$, the velocity of the center of
 mass, $V_0$, semi-major axis, $a$, the time of minima,
 $T_0$, and the orbital period, $P$ are set as free parameters.

 The light curve shows asymmetry between two maxima probably because of the presence of the stellar spots.
 We assumed a cool active region on the surface of the cooler
 component and represented it with a spot to adjust the difference between two maxima.
 The code does not give results for simultaneous solutions where the spot parameters are treated as free parameters.
In order to obtain the best parameters, we tested different solutions by changing the spot parameters.
The solution with the smallest standard deviation is regarded as the best solution for spot parameters.
Co-latitude ($\beta$), longitude ($\lambda$), fractional radius ($r$), and the
temperature factor ($t$) of the spot are given in Table~\ref{tab:V1191:lc}.
Spot parameters obtained from the LC solutions do not represent
a unique spot but the total active area on the stellar surface.
Table~\ref{tab:V1191:lc} also summarizes results of the analysis. $B$, $V$, and $R$
 light curves and velocity curves that are computed using the determined parameters are shown with solid lines in Fig.~\ref{Fig:V1191Cyg:LC}a.
 The solution presented in this study is the first simultaneous light and radial velocity data solution of the system.

\begin{table}
\caption{The light and velocity curve results and their formal 1$\sigma$ errors for \astrobj{V1191 Cyg}.
The indices $h$ and $c$ refer to the hot and cooler components, respectively. See text for details.} \label{tab:V1191:lc}
\begin{tabular}{lll}
\hline
Parameter                                   & Value      \\
\hline
$i$ ${({^\circ})}$                          & 80.1(5)   \\
$q = M_\textrm{h} / M_\textrm{c}$           & 0.105(2)   \\
$a$ ($\rm{R_{\odot}}$)			            & 2.192(72)  \\
$V_0$ (kms$^{-1}$)			                & -17.3(3) \\
$\Omega _{h}=\Omega _{c}$                   & 1.924(194)      \\
$T_\textrm{h}$ (K)                                   & 6610(200)   \\
$T_\textrm{c}$ (K)                                   & 6500  \\
Fractional radius of hot comp.              & 0.237(79) \\
Fractional radius of cooler comp.           & 0.598(70) \\
%Radiative parameters:                       &            \\
$A_\textrm{h} = A_\textrm{c}$                            & 0.6        \\
$g_\textrm{h} = g_\textrm{c}$               & 0.32        \\
%Limb darkening  $x_h, x_c$                   &            \\
%$x_1B$                                      &       \\
%$x_2B$                                      &       \\
%$x_1V$                                      &       \\
%$x_2V$                                      &       \\
%$x_1R$                                      &       \\
%$x_2R$                                      &       \\
Luminosity ratio:$\frac{L_c}{L_T}$ &   \\
$B$                                         & 0.880(50)          \\
$V$                                         & 0.878(39)        \\
$R$   					    & 0.876(20)         \\
Spot parameters:			& \\
$\beta$ ${({^\circ})}$			   & 30 \\
$\lambda$ ${({^\circ})}$			   & 290 \\
$r$ 			   		   & 20 \\
$t$				           & 0.9 \\
\hline
\end{tabular}
\end{table}

\section{Results and Conclusion}

B, V, and R light curves of the system have been solved simultaneously with the spectroscopic
study of Rucinski et al. (2008). We derived the orbital and the physical parameters of the components.
Using obtained parameters we could estimate the filling factor ($f=\frac{\Omega_{{in}}-\Omega}{\Omega_{{in}}-\Omega_{{out}}}$) from the
inner (${\Omega_{{in}}}$) to the outer critical surface (${\Omega_{{out}}}$) as 0.74.
The distance of the system to the Sun estimated as
278 pc by using observed parameters (Table~\ref{tab:V1191Cyg:par}).
This is the first distance value given for \astrobj{V1191 Cyg}.
During the calculations, effective temperature and absolute magnitude of the Sun are taken as 5780~K and 4.75 mag, respectively.
We compared our result with the well-known contact binaries using YE05 tables and figures.
The physical parameters obtained from our analysis of \astrobj{V1191 Cyg}  seem to be in a good agreement
with the well-known LTCBs. Like the other W-subtype secondaries, the less massive component of \astrobj{V1191 Cyg} is overluminous and oversized in YE05 diagrams.
While the secondary component of the system appears to be below the ZAMS, the massive and the cooler component is situated close to the TAMS.
This phenomenon was discussed in detail by YE05, Kalomeni et al. (2007),  Stepien (2006), and Gazeas \& Niarchos (2006).

Light variations at maxima and minima depths have been detected for the first time.
The light curves of Pribula et al. (2005a) indicate that the occultation minimum is
deeper than the transit by $\sim 0.01$ mag in R band.
However, light curves presented in this study (Fig.~\ref{Fig:V1191Cyg:LC}a) show that the deeper minimum is
the transit and the shallower one is the occultation.
The occultation minimum is shallower by $\sim 0.04$ mag in R band.

There are two subgroups of A- and W- types for LTCBs according to Binnendijk (1970) classification.
If the more massive component (the hotter one) is eclipsed during the deeper minimum, then we are dealing with an A-type
system. If the less massive star is the hotter one, the system is W-type.
\astrobj{FG Hya} (Qian \& Yang, 2005) and \astrobj{EM Psc (Qian} et al., 2008)
show an interchange between this two subgroups which is consistent with the model of YE05.
The light curves of Pribula et al. (2005a) show that \astrobj{V1191 Cyg} is
a W-type system whereas our observations show that the minima of the light curve have been interchanged.
Apart from the long period evolution (e.g. heat transfer between components) of LTCB systems, short period
variation (e.g. efficiently stellar activity) can affect this phenomena.
\astrobj{V1191 Cyg} shows similar characteristic properties of \astrobj{FG Hya} and \astrobj{EM Psc}.

\astrobj{V1191 Cyg} with binaries having small mass ratio and short orbital 
period form a small but important sub-group of binary zoo.
Short period and small mass ratio in contact binaries play a crucial role to explain astrophysical
phenomena that are not well understood yet (i.e. merge of close/contact binaries, blue stragglers and the formation of
\astrobj{FK Com} stars) (YE05; Arbutina, 2009; Jiang et al., 2010).
It is generally believed that these kinds of systems can be the progenitors of the \astrobj{FK Com} systems and blue stragglers.

\begin{table*}
\begin{center}
\caption{Absolute parameters of V1191 Cyg. The standard errors
1$\sigma$ in the last digit are given in parentheses. H and C stand for the hot and the cooler components, respectively}
\label{tab:V1191Cyg:par}
\begin{tabular}{llll}
\hline
Parameter                                        &Unit                      & H           & C   \\
\hline
Mass (M)                                         &$\rm{M_{\odot}}$      & $0.13(1)$            & $1.29(8)$      \\
Radius (R)                                       &$\rm{R_{\odot}}$      & $0.52(15)$           & $1.31(18)$      \\
Temperature (T$_{\rm eff}$)                      &$\rm{K}$              & $6610(200)$              & $6500$    \\
Luminosity (L)                                   &$\rm{L_{\odot}} $     & $0.46(8)$           & $2.71(44)$      \\
Surface gravity ($\log g$)                       &$\rm{cms^{-2}} $      & $4.12$             & $4.31$      \\
Absolute magnitude (M$_V$)                       &mag                   & 5.73              & 3.82           \\
Period change rate ($\dot{P}$)                   &d/yr                 &~~~~~~~~~~~~~$1.3(1)\times10^{-6}$ &      \\
Mass transfer ratio ($\dot{M}$)                  &M$_\odot$/yr         &~~~~~~~~~~~~~$2.0(4)\times10^{-7}$ &      \\
%Seperation between stars ($a$)                  &$\rm{R_{\odot}}$      &~~~~~~~~~~~~~~~~~2.192(72) &      \\
Distance (d)                                     &pc                    &~~~~~~~~~~~~~~~~~278(31) &      \\
\hline
\end{tabular}
\end{center}
\end{table*}

\section*{Acknowledgments}
This study was supported by the Turkish Scientific
and Research Council (T\"UB\.ITAK 109T047) and Ege University Research Fund.
KY+VK acknowledges support by the Turkish Academy of Sciences (T{\"U}BA).
The authors thank to E. R. Pek\"unl\"u for his valuable comments.

\end{document}